\documentclass[10pt,twocolumn,oneside,final]{IEEEtran}

\usepackage[cmex10]{amsmath}
\usepackage{amsthm}
\usepackage{amssymb}
\usepackage{amsfonts}
\usepackage{algorithm,algcompatible}
\usepackage{array}
\usepackage{mathrsfs}

\usepackage{xspace}
\usepackage{textcase}

\usepackage{wasysym}
\usepackage{graphicx}
\usepackage{subcaption}
\usepackage{color}
\usepackage{xcolor}

\usepackage{dcolumn}
\usepackage{longtable}
\usepackage{multirow}
\usepackage{booktabs}
\usepackage{tabularx}

\usepackage{times}
\usepackage{bm}

\usepackage{url}
\usepackage{cite}

\algnewcommand\INPUT{\item[\textbf{Input:}]}%
\algnewcommand\OUTPUT{\item[\textbf{Output:}]}%

\newcommand{\E}{{\mathbb{E}}} 
\newcommand{\Var}{{\mathbb{V}\mathrm{ar}}} 
\newcommand{\CN}{{\mathcal{CN}}} 
\newcommand{\RL}{{\mathbb{R}}} 
\newcommand{\CX}{{\mathbb{C}}} 
\newcommand{\tr}{{\mathrm{tr}}} 
\newcommand{\Hm}{{\mathrm{H}}} 
\newcommand{\T}{{\mathrm{T}}} 
\theoremstyle{plain}
\newtheorem{theorem}{Theorem}
\newtheorem{lemma}{Lemma}

\newtheorem{corollary}{Corollary}

\theoremstyle{definition}
\newtheorem{definition}{Definition}

\theoremstyle{remark}
\newtheorem{remark}{\bf{Remark}}

\begin{document}

\title{Framework of Channel Estimation for Hybrid Analog-and-Digital Processing Enabled Massive MIMO Communications}
\author{Leyuan Pan, \IEEEmembership{Student Member, IEEE,}
	\and Le Liang, \IEEEmembership{Student Member, IEEE,}\\
	\and Wei Xu, \IEEEmembership{Senior Member, IEEE,}
	\and and Xiaodai Dong, \IEEEmembership{Senior Member, IEEE}
	\thanks{
		L. Pan and X. Dong are with the Department of Electrical and Computer Engineering, University of Victoria, Victoria, BC V8P 5C2, Canada (Email: leyuan@uvic.ca, xdong@ece.uvic.ca).

		L. Liang is with the School of Electrical and Computer Engineering, Georgia Institute of Technology, Atlanta, GA 30332, USA (Email: lliang@gatech.edu).

		W. Xu is with the National Mobile Communications Research Laboratory, Southeast University, Nanjing, Jiangsu 210096, China (Email: wxu@seu.edu.cn).
	}
}
\maketitle

\begin{abstract}
	We investigate a general channel estimation problem in the massive multiple-input multiple-output (MIMO) system which employs the hybrid analog/digital precoding structure with limited radio-frequency (RF) chains.
	By properly designing RF combiners and performing multiple trainings, the proposed channel estimation can approach the performance of fully-digital estimations depending on the degree of channel spatial correlation and the number of RF chains.
	Dealing with the hybrid channel estimation, the optimal combiner is theoretically derived by relaxing the constant-magnitude constraint in a specific single-training scenario, which is then extended to the design of combiners for multiple trainings by \emph{Sequential} and \emph{Alternating} methods.
	Further, we develop a technique to generate the phase-only RF combiners based on the corresponding unconstrained ones to satisfy the constant-magnitude constraints.
	The performance of the proposed hybrid channel estimation scheme is examined by simulations under both nonparametric and spatial channel models.
	The simulation results demonstrate that the estimated CSI can approach the performance of fully-digital estimations in terms of both mean square error and spectral efficiency.
	Moreover, a practical spatial channel covariance estimation method is proposed and its effectiveness in hybrid channel estimation is verified by simulations.
\end{abstract}


\section{Introduction}\label{chpt:HCE:sec:Introduction}
\IEEEPARstart{M}{assive} multiple-input multiple-output (MIMO) is potentially one of the most promising and key technologies to meet the stringent performance requirements for the next-generation, i.e., 5G, wireless communications \cite{Andrews2014What,Chin2014Emerging}.
It has attracted considerable research interests from both academia and industry since the seminal work \cite{Marzetta2010Noncooperative} was published in 2010 \cite{Hoydis2013Massive,Larsson2014Massive,Jin2014Zero,Dai2016Power}.
Most notably, with an excessive amount of antennas mounted at the transmitter and/or receiver, signal processing, including both transmission precoding and receiving combining, can be greatly simplified while achieving highly optimal performance \cite{Pan2016novel,Rusek2013Scaling,Bjoernson2015Optimal,Pan2017Multipair}.
Simple linear precoding schemes, such as zero-forcing (ZF), are virtually optimal and comparable to the capacity-achieving nonlinear dirty paper coding (DPC).
Thanks to the employment of millimeter wave (mmWave) bands, a large number of antennas can be packed into a small area due to the short wavelength \cite{HeathJr.2016overview}.

However, the conventional signal processing is typically performed at the baseband, which means that the signal received from each antenna port needs to be properly filtered, down-converted, and then sampled, where the hardware module performing such tasks is normally referred to as a radio-frequency (RF) chain.
Analogous procedure exists for signal transmission.
Given a large number of antennas in a massive MIMO system, it would be formidable to feed each antenna with a dedicated RF chain due to high cost and power consumption.
To circumvent the challenging requirement of massive RF chains, an analog/digital hybrid structure has been proposed for massive MIMO systems operated on both centimeter wave (cmWave) \cite{Zhang2005Variable} and mmWave \cite{Ayach2012capacity} bands\footnote{According to the notation in \cite{Molisch2016Hybrid}, the cmWave bands generally denote $1-10$ GHz while mmWave denotes $10-100$ GHz.}.
On the transmitter side of a hybrid system, low-dimensional baseband signals (after digital processing) are converted to the RF domain, feeding a phase-shifting network to properly adjust the phases of transmission signals which are then transmitted to wireless channels by antennas \cite{Zhang2005Variable,HeathJr.2016overview}.
The design of the RF and baseband precoding/combining matrices in a hybrid structure is a challenging problem and has been extensively studied in recent years, e.g., \cite{Ayach2012capacity,Ayach2014Spatially,Liang2014Low,Alkhateeb2015Limited,Alkhateeb2016Frequency}.
However, these papers all assume the availability of channel state information (CSI).
Other beamforming solutions without the explicit need of channel knowledge depending on iterative beam trainings and multi-stage codebooks have been developed in \cite{Wang2009Beam,Chen2011Multi, Hur2013Millimeter,Tsang2011Coding}.
These solutions, however, naturally have the common disadvantage that the beamforming converges towards a single beam which is not capable to achieve multiplexing gains or support multiple streams in multi-user systems.
Hence, the explicit channel estimation is one of the most important elements in reaping all the advantages and gains of massive MIMO, and unsurprisingly has been under extensive investigation.
It is a challenging task, yet is even more difficult to fulfill under the limited RF-chain hybrid structure constraint.

The main task of the channel estimation in a hybrid precoding system is to recover the $M$-dimensional channel vector from $L$ observations at baseband, where $L$ ($< M$) is the number of limited RF chains.
The conventional fully-digital MIMO channel estimation methods in the literature cannot be applied in a hybrid structure to obtain full CSI \cite{Choi2014Downlink,Biguesh2006Training,Wen2015Channel,Fan2015Channel,Xie2016ULDL}.
So far, only a few papers have studied channel estimation with limited RF chains in hybrid structured mmWave communications.
An adaptive compressed sensing solution for the hybrid channel estimation was proposed in \cite{Alkhateeb2014Channel} from the perspective of angular sparsity of mmWave channels.
Scanning in the angular domain is performed at both the transmitter and receiver sides.
As such, the complexity and resource consumption of the proposed channel estimation scheme is dominated by the sparsity of channel scattering.
For instance, it will consume much more resource with higher complexity to achieve a desired angular resolution in channel estimation when the scattering paths are rich, compared to the case of very sparse channels.
A similar method was developed in a frequency-selective channel scenario \cite{Gao2016Channel}.
In \cite{Mendez-Rial2015Channel}, channel estimation with phase-shifters or switches in a hybrid structured system was discussed.
Still, it employs the high-complexity compressed sensing method with angular domain scanning to estimate the angle of arrival/angle of departure (AoA/AoD) of each scattering path.
Few prior literature has considered the general problem of channel estimation in a hybrid structured massive MIMO system without significant dependence on the channel sparsity.
Since the degree of freedom (DoF) of the received signals at baseband is limited by the number of RF chains, it becomes prohibitively difficult to obtain satisfactory higher-dimensional vector channel estimates, especially for rich scattering environments, in a hybrid precoding system using conventional channel estimation approaches.

In this paper, we consider the uplink channel estimation of a multi-user massive MIMO system with the hybrid RF-baseband processing structure where the base station (BS) is equipped with a large number of antennas but driven by a far smaller number of RF chains, and each mobile station (MS) is equipped with a single antenna, and propose an efficient channel estimation scheme by exploiting the spatial correlation of massive MIMO channels where both single and multiple trainings are investigated.
One may suspect the necessity of investigation on multiple trainings.
Theoretically, one pilot symbol is optimal, and in fact enough, for a single-antenna user to assist uplink estimation of uncorrelated MIMO channels under some mild assumptions \cite{Hassibi2003How}.
In a massive MIMO system with limited RF chains, however, only much smaller $L$-dimensional (compared to the number of antennas, $M$) signal is captured at baseband per training.
It is, therefore, evident that a single training symbol becomes insufficient for the BS to conduct full-dimensional channel estimation of $M$ (independent) channel coefficients.
To approach the performance of fully-digital channel estimation, multiple trainings are required to achieve the DoF of fully-digital baseband signal measurements.
That is, $TL$ observations can be utilized to estimate the $M$-dimensional MIMO channels, where $T$ is the number of trainings.
Hence, it needs to properly design RF combiners for different training phases to capture the channel energy and then recover the CSI as accurately as possible.
Furthermore, how many trainings are required to achieve the performance of fully-digital channel estimation?
Empirically, $T=M/L$ training symbols can be utilized to achieve the DoF of a fully-digital training in the estimation of uncorrelated channels.
However, such conclusions may not hold for correlated channels.
As revealed in both \cite{Kotecha2004Transmit} and \cite{Bjoernson2010framework}, the optimal number of training symbols can be reduced due to the fact that the dimension of statistically dominant subspaces is less than the number of antennas.
In practice, massive MIMO and mmWave channels are inevitably spatially correlated due to the limited number of propagation paths and angular spreads \cite{Alkhateeb2014Channel,Alkhateebt2014Single,Choi2014Downlink}.
Thus, less than $M/L$ trainings could be sufficient to achieve the fully-digital channel estimation performance in such spatially correlated channels.
Finally, we summarize the main contributions of this paper as follows:
\begin{itemize}
	\item \textbf{RF Combiner Design for Single Training:}
	We investigate the channel estimation in a hybrid precoding system and formulate an optimization problem on channel estimation following the minimum mean square error (MMSE) criteria.
	To solve the problem, the constant-magnitude constraint of the RF combiner is temporarily relaxed.
	By employing the properties of \emph{Block Generalized Rayleigh Quotient}, the theoretical optimizer of the formulated optimization problem is solved and the corresponding optimal RF combiner is designed in the single training scenario.
	Analyses of the designed combiner show that the mean square error (MSE) of the channel estimation decreases when \emph{(1) the channel is more spatially correlated, (2) more RF chains for channel estimation are deployed, (3) larger transmission power for training pilots is utilized}.
	Moreover, the \emph{closed-form} expression of the MSE is derived and verified to be quite accurate compared to the simulation results, thus providing useful guidelines in practical system designs.
	\item \textbf{RF Combiner Design for Multiple Trainings:}
	We formulate the optimization problem to design RF combiners for multiple trainings and solve it by \emph{Sequential} and \emph{Alternating} methods with the assistance of the single-training result.
	The \emph{Sequential} method can achieve step-wisely minimum MSE with low complexity while the \emph{Alternating} one solves the joint optimization problem iteratively with high complexity, however, achieving the local optimum.
	The performance and complexity of the proposed RF combiners are examined in simulations under both nonparametric and parametric channel models \cite{Forenza2007Simplified}.
	\item \textbf{Spatial Channel Covariance Estimation in Hybrid-Structured System:}
	To perform the proposed RF combiner design and channel estimation, the spatial channel covariance matrix needs be known by the BS to perform the design of RF combiners for channel estimation.
	In this paper, a covariance estimation method is proposed in the hybrid-structured massive MIMO system.
	From simulation results, it achieves comparable spectral efficiency with the estimated covariance matrix compared to that with the perfect one, which proves the effectiveness of the proposed method.
\end{itemize}
Note that the considered system structure and channel estimation scheme is also applicable to mmWave channels \cite{Molisch2016Hybrid,Liang2014Low,Bogale2016Number}.

Apart from the contributions, we also present the main characteristics differentiating this paper from the existing works as summarized below.
\begin{itemize}
	\item Compared to the typical fully-digital massive MIMO systems, the signal dimension in the considered system is reduced after the processing of the phase-shifting network.
	That is, it converts the received $M$-dimensional RF signal to $L$-dimensional baseband signal where $M$ denotes the number of antenna elements and $L$ represents the number of RF chains typically satisfying $L<M$ in hybrid precoding/combining systems.
	As such, the channel estimation performed in baseband cannot utilize the complete training information as the fully-digital massive MIMO does.
	\item Compared to existing work on channel estimation in a hybrid-structure mmWave system, no explicit usage of channel sparsity is required by our proposed methods to perform the estimation, which is adaptive to both non-sparse cmWave and sparse mmWave channel scenarios.
	In addition, the complexity of our proposed scheme is determined by the number of RF chains and training phases which are fixed in an online system, rather than the scattering circumstance which is time-variant and difficult to be guaranteed.
\end{itemize}

\emph{Organization:}
The rest of this paper is organized as follows.
Section \ref{chpt:HCE:sec:SystemModel} illustrates the system model of the massive MIMO system with the hybrid precoding structure and the channel models adopted in this paper.
In Section \ref{chpt:HCE:sec:ChannelEstimationWithHybridStructure}, the channel estimation schemes in the hybrid structure are proposed and analyzed.
Section \ref{chpt:HCE:sec:EstimateChannelCorrelationsByCovarianceMatching} proposes the method to generate the channel covariance matrices which can be employed in practical system implementations.
The numerical and simulation results are presented in Section \ref{chpt:HCE:sec:NumericalSimulationResults}.
Finally, Section \ref{chpt:HCE:sec:Conclusions} concludes our work in this paper.

\emph{Notation:}
The boldface capital and lowercase letters are used to denote matrices and vectors, respectively, while the plain lowercase letters are scalars.
Superscripts $(\cdot)^*$, $(\cdot)^\T$ and $(\cdot)^\Hm$ stand for the conjugate, transpose and conjugate-transpose of a vector or matrix, respectively.
$\mathbf{I}_M$ represents the identity matrix of size $M$.
Operators $\|\cdot\|$, $\|\cdot\|_\mathrm{F}$ and $\mathrm{tr}(\cdot)$ denote the Euclidean norm of a vector, the Frobenius norm and the trace of a matrix, respectively.
$\mathrm{blkdiag}\{\mathbf{A}_1, \cdots, \mathbf{A}_L\}$ stands for the block matrix with diagonal elements as $\mathbf{A}_1,\cdots,\mathbf{A}_L$ in sequence.
For statistical vectors and matrices, $\E\{\cdot\}$ and $\Var\{\cdot\}$ are utilized to represent the expectation and variance, respectively.
Moreover, $\mathbf{x}\sim\CN(\mathbf{0}, \bm{\Sigma})$ represents the complex Gaussian random vector $\mathbf{x}$ with zero mean and covariance matrix $\bm{\Sigma}$.
Finally, $\mathbf{a}\succ\mathbf{b}$ means that $\mathbf{a}$ majorizes (or dominates) $\mathbf{b}$ and vice versa.

\section{System Model}\label{chpt:HCE:sec:SystemModel}
\begin{figure}[htbp]
	\centering
	\includegraphics[width=0.9\linewidth]{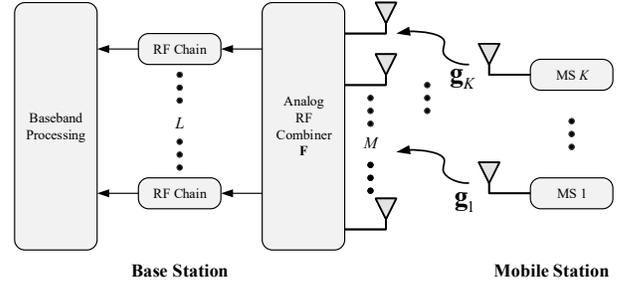}
	\caption{Block diagram of massive MIMO with a hybrid structure.}
	\label{chpt:HCE:fig:SystemModel}
\end{figure}
We investigate a single-cell hybrid massive MIMO communication structure as shown in Fig. \ref{chpt:HCE:fig:SystemModel} where $K$ single-antenna mobile stations (MSs) are served by a base station (BS) equipped with $M$ antennas driven by $L$ RF chains which are interfaced by an analog RF phase-shifting network converting the high-dimensional received signal to low-dimensional one through phase-only linear combing.
Here, $L<M$ means limited RF chains while $L=M$ implies full chains.
The combined RF signal after the phase-shifting network is then down-converted to baseband and sampled by analog-to-digital converter(s) for further digital processing \cite{HeathJr.2016overview}.

Due to its reciprocity in TDD operation, the channel between the BS and each MS can be obtained by the uplink training and used for downlink precoding within a coherence interval \cite{Marzetta2006How}.
Different from the downlink training employed in \cite{Alkhateeb2015Compressed}, the uplink training requires the pilot sequences transmitted by all MSs to be orthogonal to each other to avoid pilot contamination.
Hence, the duration of each pilot sequence should be at least $K$ symbols to guarantee the orthogonality conditions \cite{Marzetta2006How}.
Among all, the identity matrix $\mathbf{I}_K$ is one of the simplest orthogonal pilots, which implies that the MSs transmit pilot sequentially and all others keep mute when a MS is transmitting pilot.
Moreover, the performance of channel estimation with identity matrix as pilot is the same as that with others under block-fading channels.
On the other hand, it is more practical to design the channel-estimation-used analog combiner based on the corresponding spatial channel covariance matrix of a single MS per training, which can be understood from the perspective of hybrid beamforming that the beamformer is toward one of the uncorrelated MSs when it is transmitting pilot signal and leads to better performance in sparse mmWave channels where the number of scattering paths is limited.
Consequently, the channel estimation of each MS can be performed one by one with similar manipulations.
For simplicity, we select a single MS to illustrate the hybrid channel estimation framework in the following elaborations.

During the uplink training period, a single-antenna MS transmits $T$ pilots to the BS.
Considering the $t$th ($t=1,2,\cdots,T$) pilot training, the received baseband signal at BS after pilot compensation can be expressed as
\begin{equation}
	\mathbf{y}_t = \mathbf{F}_t\left(\sqrt{\rho}\mathbf{g}\varphi_t + \mathbf{\tilde{n}}_t\right)\varphi_t^* = \sqrt{\rho}\mathbf{F}_t\mathbf{g} + \mathbf{F}_t\mathbf{n}_t,
\end{equation}
where $\mathbf{g}\in\CX^{M\times1}$ represents the uplink channel from the MS to BS which is assumed to remain static throughout a coherence interval, $\varphi_t$ stands for the $t\text{th}$ training symbol with $\varphi_t\varphi_t^*=1$, $\rho$ is the pilot power per transmission, $\mathbf{\tilde{n}}_t\in\CX^{M\times1}$ denotes the additive white Gaussian noise (AWGN) with zero mean and unit variance, $\mathbf{n}_t\triangleq\mathbf{\tilde{n}}_t\varphi_t^*$ which persists as $\CN(\mathbf{0}, \mathbf{I}_M)$ AWGN, and $\mathbf{F}_t\in\CX^{L\times{M}}$ is the RF phase-shifting matrix constructed by unit-magnitude elements.
For $T$ pilot transmissions, the received signal can be stacked as a concatenated vector $\mathbf{y}_\mathrm{c} = [\mathbf{y}_1^\T, \mathbf{y}_2^\T, \cdots, \mathbf{y}_T^\T]^\T\in\CX^{TL\times1}$ given by
\begin{equation}\label{chpt:HCE:eq:CompactInputOutput}
	\mathbf{y}_\mathrm{c} = \sqrt{\rho}\mathbf{F}_\mathrm{c}\mathbf{g} + \mathbf{F}_\mathrm{d}\mathbf{n}_\mathrm{c},
\end{equation}
where the pilot power is identical for each training period, the compact matrices $\mathbf{F}_\mathrm{c} = [\mathbf{F}_1^\T, \cdots, \mathbf{F}_T^\T]^\T$, $\mathbf{F}_\mathrm{d} = \mathrm{blkdiag}\{\mathbf{F}_1, \cdots, \mathbf{F}_T\}$, and $\mathbf{n}_\mathrm{c} = [\mathbf{n}_1^\T, \cdots, \mathbf{n}_T^\T]^\T$.

In this paper, we consider the spatially correlated MIMO channel model, which is a typical case in most MIMO scenarios due to the limited number of incident paths and angular spreads on BS \cite{Choi2014Downlink}.
The spatial channel covariance matrix can be expressed by $\mathbf{R}=\E\{\mathbf{g}\mathbf{g}^\mathrm{H}\}$.
Without loss of generality, we normalize $\tr\left(\mathbf{R}\right)=M$. 
In Section \ref{chpt:HCE:sec:ChannelEstimationWithHybridStructure}, $\mathbf{R}$ is assumed to be known by the BS while the estimation of the covariance matrix in practice will be discussed in Section \ref{chpt:HCE:sec:EstimationOfSpatialChannelCovariance}.

\section{Channel Estimation with Hybrid Structure}\label{chpt:HCE:sec:ChannelEstimationWithHybridStructure}
For channel estimation at the BS, the task is to obtain the channel estimate in the MMSE sense with the designed phase-shifting matrix.
By employing the MMSE aimed linear estimator, the channel estimation problem can be formulated as follows:
\begin{equation}\label{chpt:HCE:opt:JointChannEst}
	\begin{aligned}
		&\underset{\mathbf{F}_\mathrm{c}, \mathbf{W}}{\text{minimize}} && \E\left\{\|\mathbf{g}-\mathbf{\hat{g}}\|^2\right\} \\
		&\text{subject to:} && \mathbf{\hat{g}}=\mathbf{W}\mathbf{y}_\mathrm{c}\\
		&&& \mathbf{y}_\mathrm{c} = \sqrt{\rho}\mathbf{F}_\mathrm{c}\mathbf{g} + \mathbf{F}_\mathrm{d}\mathbf{n}_\mathrm{c} \\
		&&& \mathbf{F}_i\in\mathcal{F}, i=1,\cdots,T
	\end{aligned}
\end{equation}
where $\mathbf{W}$ represents the baseband processing matrix, and $\mathcal{F}$ denotes the set of all feasible phase-only RF combiners.
In order to minimize MSE, the optimal solution to $\mathbf{W}$ is the well-known Weinner filter \cite{Kay1993Fundamentals} given by
\begin{equation}\label{chpt:HCE:opt:WeinnerFilter}
	\mathbf{W}=\sqrt{\rho}\mathbf{R}\mathbf{F}_\mathrm{c}^\Hm\left(\rho\mathbf{F}_\mathrm{c}\mathbf{R}\mathbf{F}_\mathrm{c}^\Hm+\mathbf{R}_{\mathbf{F}_\mathrm{d}}\right)^{-1},
\end{equation}
where $\mathbf{F}_\mathrm{c}$ is assumed to be known as a prior and $\mathbf{R}_{\mathbf{F}_\mathrm{d}}\triangleq\mathbf{F}_\mathrm{d}\mathbf{F}_\mathrm{d}^\Hm$ for notation simplification.
Using (\ref{chpt:HCE:opt:WeinnerFilter}), the objective function in (\ref{chpt:HCE:opt:JointChannEst}) equals
\begin{align}
	\mathrm{MSE} = & \E\left\{\left\|\mathbf{g}-\mathbf{\hat{g}}\right\|^2\right\} \nonumber\\
	= & \tr\left(\mathbf{R}-\rho\mathbf{R}\mathbf{F}_\mathrm{c}^\Hm\left(\rho\mathbf{F}_\mathrm{c}\mathbf{R}\mathbf{F}_\mathrm{c}^\Hm+\mathbf{R}_{\mathbf{F}_\mathrm{d}}\right)^{-1}\mathbf{F}_\mathrm{c}\mathbf{R}\right) \nonumber \\
	= & \tr\left(\left(\mathbf{R}^{-1}+\rho\mathbf{F}_\mathrm{c}^\Hm\mathbf{R}_{\mathbf{F}_\mathrm{d}}^{-1}\mathbf{F}_\mathrm{c}\right)^{-1}\right) \label{chpt:HCE:eq:ChannEstMSE},
\end{align}
where the last equality holds due to the Woodbury matrix identity \cite{Higham2002Accuracy} and it is assumed that $\mathbf{R}$ is in full rank.
If $\mathbf{R}$ is rank deficient, remedies similar to the one discussed in \cite{Kotecha2004Transmit} can be applied which does not affect the MSE.
Combining (\ref{chpt:HCE:opt:JointChannEst})--(\ref{chpt:HCE:eq:ChannEstMSE}), the primal channel estimation problem is readily equivalent to the following problem as
\begin{equation}\label{chpt:HCE:opt:DesignPhaseShifter}
	\begin{aligned}
		&\underset{\mathbf{F}_\mathrm{c}}{\text{minimize}} && \tr\left(\left(\mathbf{R}^{-1}+\rho\mathbf{F}_\mathrm{c}^\Hm\mathbf{R}_{\mathbf{F}_\mathrm{d}}^{-1}\mathbf{F}_\mathrm{c}\right)^{-1}\right) \\
		&\text{subject to:} && \mathbf{F}_t\in\mathcal{F}, t=1,\cdots,T.
	\end{aligned}
\end{equation}
The optimization problem (\ref{chpt:HCE:opt:DesignPhaseShifter}) aims to find the optimal and feasible RF combiners $\mathbf{F}_t$ ($t=1,\cdots,T$) for the $T$-length training sequence.
However, it is difficult to directly obtain the optimal solution in either closed-form expressions or through numerical approaches without sustainable resource consumption due to the non-convex constraint.

\emph{To facilitate further analysis, we temporarily drop the constant-magnitude constraint in the following design.}
This is a good way to begin with because it is capable to derive some closed-form expressions and provide guidelines for the design of phase-only combiners.
Note that the design of phase-only combiners will be considered in Section \ref{chpt:HCE:sec:DesignOfPhaseOnlyCombiners}.

Without the phase-only constraint, it can be shown that the optimal RF unconstrained combiner has a structure characterized in the following lemma.
\begin{lemma}\label{chpt:HCE:lem:UnconstrainedRFCombinerStructure}
	For any optimal solution to the unconstrained RF combiners given their singular value decompositions (SVD) as 
	\begin{equation}\label{chpt:HCE:eq:UnconstrainedRFCombinerSVD}
		\tilde{\mathbf{F}}_t^{\mathrm{opt}} = \mathbf{U}_t[\bm{\Sigma}_t~\mathbf{0}][\mathbf{V}_{t,\mathrm{L}}~\mathbf{V}_{t,\mathrm{R}}]^\mathrm{H},
	\end{equation}
	it is safe to reconstruct the optimal combiners according to 
	\begin{equation}\label{chpt:HCE:eq:UnconstrainedRFCombinerStructure}
		\mathbf{F}^\mathrm{opt}_t = \mathbf{V}_{t,\mathrm{L}}^\mathrm{H},
	\end{equation}
	which yields the same MSE performance as $\tilde{\mathbf{F}}_t^{\mathrm{opt}}$ does, where $t=1,\cdots,T$
\end{lemma}
\begin{IEEEproof}
	See Appendix \ref{chpt:HCE:prf:lem:UnconstrainedRFCombinerStructure}
\end{IEEEproof}

Lemma \ref{chpt:HCE:lem:UnconstrainedRFCombinerStructure} indicates that the optimal unconstrained RF combiner $\mathbf{F}^\mathrm{opt}_t$ can always be restricted as row-unitary and is thus \emph{independent} of both $\mathbf{U}_t$ and $\bm\Sigma_t$.
Hence, we can set both of them to be identity matrices and equivalently simplify the MSE in (\ref{chpt:HCE:eq:ChannEstMSE}) as
\begin{equation}\label{chpt:HCE:eq:ModifiedGeneralMSE}
	\mathrm{MSE} = \tr\left(\left(\bm{\Lambda}^{-1} + \rho\sum_{t=1}^{T}\mathbf{\tilde{V}}_t\mathbf{\tilde{V}}_t^\Hm\right)^{-1}\right),
\end{equation}
where $\mathbf{\tilde{V}}_t\triangleq\mathbf{U}^\Hm\mathbf{V}_{t,\mathrm{L}}$, $\bm{\Lambda}=\mathrm{diag}\{\lambda_1, \cdots, \lambda_M\}$ and $\mathbf{U}$ are the eigenvalues and corresponding eigenvectors of $\mathbf{R}$, i.e., $\mathbf{R}=\mathbf{U}\bm{\Lambda}\mathbf{U}^\Hm$, respectively.
Without loss of generality, it is assumed that the eigenvalues are arranged in the decreasing order, i.e., $\lambda_1\geq\cdots\geq\lambda_M$, and hence $\sum_{t=1}^{M}\lambda_t=\tr(\mathbf{R})=M$.
In addition, there exists $\mathbf{\tilde{V}}_t^\Hm\mathbf{\tilde{V}}_t=\mathbf{I}_L$ due to the unitary property of both $\mathbf{U}$ and $\mathbf{V}_{t,\mathrm{L}}$.
To design the optimal RF combiners, we first investigate the single-pilot strategy, i.e., $T=1$, and then extend to the multiple pilot trainings where $T>1$.

\subsection{Optimal Combiner Design of Single Training}\label{chpt:HCE:sec:OptimalCombinerDesignSinglePilotTraining}
When $T=1$, the MSE expressed in (\ref{chpt:HCE:eq:ModifiedGeneralMSE}) reduces to
\begin{subequations}\label{chpt:HCE:eq:ModifiedSingleShotMSE}
	\begin{align}
		\mathrm{MSE} &= \tr\left(\bm{\Lambda}-\bm{\Lambda}\mathbf{\mathbf{\tilde{V}}}_1\left(\mathbf{\tilde{V}}_1^\Hm\bm{\Lambda}\mathbf{\tilde{V}}_1+\rho^{-1}\mathbf{I}_M\right)^{-1}\mathbf{\tilde{V}}_1^\Hm\bm{\Lambda}\right) \label{chpt:HCE:eq:ModifiedSingleShotMSE2} \\
		&= \tr(\bm{\Lambda}) - \tr\left(\left(\mathbf{\tilde{V}}_1^\Hm\left(\bm{\Lambda}+\rho^{-1}\mathbf{I}_M\right)\mathbf{\tilde{V}}_1\right)^{-1}\mathbf{\tilde{V}}_1^\Hm\bm{\Lambda}^2\mathbf{\tilde{V}}_1\right), \label{chpt:HCE:eq:ModifiedSingleShotMSE3}
	\end{align}
\end{subequations}
where (\ref{chpt:HCE:eq:ModifiedSingleShotMSE2}) follows from the Woodbury matrix identity \cite{Higham2002Accuracy} and (\ref{chpt:HCE:eq:ModifiedSingleShotMSE3}) is derived on the fact that $\tr(\mathbf{AB})=\tr(\mathbf{BA})$.
From (\ref{chpt:HCE:eq:ModifiedSingleShotMSE}), the design of the unconstrained optimal RF combiner in (\ref{chpt:HCE:opt:DesignPhaseShifter}) can be reformulated as the following optimization problem:
\begin{equation}\label{chpt:HCE:opt:DesignUnconstrainedRFCombinerSingleTraining}
	\begin{aligned}
		&\underset{\mathbf{\tilde{V}}_1}{\text{maximize}}\quad\tr\left(\left(\mathbf{\tilde{V}}_1^\Hm\left(\bm{\Lambda}+\rho^{-1}\mathbf{I}_M\right)\mathbf{\tilde{V}}_1\right)^{-1}\mathbf{\tilde{V}}_1^\Hm\bm{\Lambda}^2\mathbf{\tilde{V}}_1\right) \\
		&\text{subject to:}\quad\mathbf{\tilde{V}}_1^\Hm\mathbf{\tilde{V}}_1=\mathbf{I}_L.
	\end{aligned}
\end{equation}

To solve (\ref{chpt:HCE:opt:DesignUnconstrainedRFCombinerSingleTraining}), we employ the \emph{block generalized Rayleigh quotient} which is presented as Lemma \ref{chpt:HCE:lm:BlockGeneralizedRayleighQuotient} in Appendix \ref{chpt:HCE:apdx:BlockGeneralizedRayleighQuotient}.
Applying Lemma \ref{chpt:HCE:lm:BlockGeneralizedRayleighQuotient}, the objective function of (\ref{chpt:HCE:opt:DesignUnconstrainedRFCombinerSingleTraining}) is the \emph{block generalized Rayleigh quotient} of $\mathbf{\tilde{V}}_1$ with respect to the pencil $(\mathbf{\Lambda}^2, \bm{\Lambda}+\rho^{-1}\mathbf{I}_M)$.
The property of \emph{block generalized Rayleigh quotient} described by Lemma \ref{chpt:HCE:lm:BlockGeneralizedRayleighQuotient} indicates that the maximizer of (\ref{chpt:HCE:opt:DesignUnconstrainedRFCombinerSingleTraining}) is the matrix spanned by the $L$ generalized eigenvectors corresponding to the $L$ largest eigenvalues of the pencil $(\mathbf{\Lambda}^2, \bm{\Lambda}+\rho^{-1}\mathbf{I}_M)$ if the matrix happens to satisfy the constraint in (\ref{chpt:HCE:opt:DesignUnconstrainedRFCombinerSingleTraining}).
As $\mathbf{\Lambda}^2$ and $\bm{\Lambda}+\rho^{-1}\mathbf{I}_M$ are both diagonal, the generalized eigenvalues and corresponding eigenvectors can be directly obtained as $\bm{\tilde{\Lambda}} = \left[\tilde{\lambda}_1, \cdots, \tilde{\lambda}_M\right]$ and $\mathbf{\tilde{U}}=\mathbf{I}_M$, respectively, where $\tilde{\lambda}_i = \lambda_i^2/(\lambda_i+\rho^{-1})$ ($i=1,\cdots,M$) and $\tilde{\lambda}_1\geq\cdots\geq\tilde{\lambda}_M$ according to the decreasing order of $\lambda_i$.
Therefore, the maximizer of (\ref{chpt:HCE:opt:DesignUnconstrainedRFCombinerSingleTraining}) is constructed by the eigenvectors corresponding to the largest $L$ eigenvalues of the pencil $(\mathbf{\Lambda}^2, \bm{\Lambda}+\rho^{-1}\mathbf{I}_M)$, denoted as
\begin{equation}\label{chpt:HCE:eq:UnconstrainedOptimizerTildeV}
	\mathbf{\tilde{V}}^\mathrm{opt}_1 = \mathbf{\tilde{U}}_{[1:L]} = \mathbf{I}_{M[1:L]},
\end{equation}
where $(\mathbf{\tilde{V}}^\mathrm{opt}_1)^\mathrm{H}\mathbf{\tilde{V}}^\mathrm{opt}_1=\mathbf{I}_L$ satisfies the constraint in (\ref{chpt:HCE:opt:DesignUnconstrainedRFCombinerSingleTraining}).
Now, we can arrive at the following Theorem \ref{chpt:HCE:thm:UnconstrainedOptimalRFCombiner} which presents the optimal design of the unconstrained single-training RF combiner.

\begin{theorem}\label{chpt:HCE:thm:UnconstrainedOptimalRFCombiner}
	The optimal unconstrained RF combiner for the single pilot strategy can be designed as follows:
	\begin{equation}\label{chpt:HCE:eq:UnconstrainedOptimizerRFCombiner}
		\mathbf{F}^\mathrm{opt}_1 = (\mathbf{U}_{[1:L]})^\Hm,
	\end{equation}
	where $\mathbf{U}$ is the matrix spanned by the eigenvectors of $\mathbf{R}$.
\end{theorem}
\begin{IEEEproof}
	By substituting (\ref{chpt:HCE:eq:UnconstrainedOptimizerTildeV}) into (\ref{chpt:HCE:eq:UnconstrainedRFCombinerStructure}) with the transition defined by $\mathbf{\tilde{V}}_t\triangleq\mathbf{U}^\Hm\mathbf{V}_{t,\mathrm{L}}$, the optimal unconstrained single-training RF combiner can be obtained.
\end{IEEEproof}

Theorem \ref{chpt:HCE:thm:UnconstrainedOptimalRFCombiner} implies that the RF combiner should receive the training pilots along the largest $L$ dominant eigen-directions of $\mathbf{R}$ to minimize the MSE of channel estimations.
More explicitly, we substitute (\ref{chpt:HCE:eq:UnconstrainedOptimizerTildeV}) into (\ref{chpt:HCE:eq:ModifiedSingleShotMSE}) and obtain the optimal MSE as follows:
\begin{align}
	\mathrm{MSE} =& \tr\left(\left(\bm{\Lambda}^{-1}+\rho\begin{bmatrix}
		\mathbf{I}_L & \\
		& \mathbf{0}_{M-L}
	\end{bmatrix}\right)^{-1}\right) \nonumber\\
	=& \sum_{l=1}^{L}\frac{\lambda_l}{1+\rho\lambda_l} + \sum_{l=L+1}^{M}\lambda_l \nonumber\\
	=& M - \sum_{l=1}^{L}\frac{\lambda_l^2}{\lambda_l + 1/\rho}, \label{chpt:HCE:eq:MSEOfSingleTraining}
\end{align}
where (\ref{chpt:HCE:eq:MSEOfSingleTraining}) is obtained by applying the power constraint defined by $\sum_{i=1}^{M}\lambda_i=M$.
Prior to stating more insights obtained from (\ref{chpt:HCE:eq:MSEOfSingleTraining}), we define a useful concept of one channel being \emph{more spatially correlated} than another in the following way.
\begin{definition}\label{chpt:HCE:dfn:MoreSpatiallyCorrelated}
	Let channel vectors $\mathbf{g}_1\in\CX^M$ and $\mathbf{g}_2\in\CX^M$ have covariance matrices $\mathbf{R}_1\in\RL^{M\times M}$ and $\mathbf{R}_2\in\RL^{M\times M}$, respectively.
	We say $\mathbf{g}_1$ is \emph{more spatially correlated} than $\mathbf{g}_2$ if and only if $\bm{\lambda}_1\succ\bm{\lambda}_2$, where $\bm{\lambda}_1$ and $\bm{\lambda}_2$ are composed of the eigenvalues sorted in a descending order of $\mathbf{R}_1$ and $\mathbf{R}_2$, respectively.
\end{definition}

By Definition \ref{chpt:HCE:dfn:MoreSpatiallyCorrelated} and observing (\ref{chpt:HCE:eq:MSEOfSingleTraining}), we can conclude the following corollaries on the design insights of the optimal unconstrained RF combiner revealed in Theorem \ref{chpt:HCE:thm:UnconstrainedOptimalRFCombiner}.

\begin{corollary}\label{chpt:HCE:cry:UnconstrainedOptimizerRFCombiner1}
	Given the number of RF chains and pilot power fixed, the MSE of channel estimation with the optimal unconstrained single-training RF combiner decreases if the channel is more spatially correlated.
\end{corollary}
\begin{IEEEproof}
	See Appendix \ref{chpt:HCE:prf:cry:UnconstrainedOptimizerRFCombiner1}.
\end{IEEEproof}

\begin{corollary}\label{chpt:HCE:cry:UnconstrainedOptimizerRFCombiner2}
	Given the channel correlation and pilot power fixed, the MSE of channel estimation with the optimal unconstrained single-training RF combiner decreases with more RF chains deployed.
\end{corollary}

\begin{corollary}\label{chpt:HCE:cry:UnconstrainedOptimizerRFCombiner3}
	Given the channel correlation and number of RF chains fixed, the MSE of channel estimation with the optimal unconstrained single-training RF combiner decreases with increasing pilot power.
\end{corollary}

Corollary \ref{chpt:HCE:cry:UnconstrainedOptimizerRFCombiner2} and \ref{chpt:HCE:cry:UnconstrainedOptimizerRFCombiner3} can be directly proved by checking the fact that the MSE in (\ref{chpt:HCE:eq:MSEOfSingleTraining}) monotonously decreases with the increase of $\rho$ and/or $L$.

\subsection{Combiner Design of Multiple Trainings}
In a typical fully-digital MIMO system, the $M$-dimensional signal can be observed at baseband to estimate the $M$-dimensional MIMO channels.
However, the hybrid structure MIMO system can only capture $L$-dimensional signal from the phase-shifting network fed by $M$ antennas for each training.
Note that $L<M$ in the limited RF chain scenario.
Hence, the observed low-dimensional signal is not sufficient to recover the high-dimensional channel information for a single training.
To approach the performance of fully-digital channel estimation, multiple trainings can be employed to achieve the DoF of fully-digital baseband signal measurements, namely, $TL$ observations can be utilized to estimate the $M$-dimensional MIMO channels.
Typically, it is assumed that $TL\leq M$.
Therefore, the RF combiner for each training phase needs to be properly designed to capture the channel energy for channel recovery as accurate as possible. 

In this subsection, we investigate the design of unconstrained combiners for multiple trainings, i.e., $\mathbf{F}_1, \cdots, \mathbf{F}_T$.
Intuitively, the multiple trainings can be performed by simply repeating the single-pilot training for multiple times.
Inspired by Theorem \ref{chpt:HCE:thm:UnconstrainedOptimalRFCombiner} where the most dominated $L$ eigen-directions of $\mathbf{R}$ are selected to construct the combiner, it is heuristic to select the most dominated $TL$ eigen-directions to establish the $T$ combiners for multiple trainings, namely, the \emph{Block Selection} method, where the RF combiner for the $t$th ($t=1,2,\cdots,T$) training can be composed of the $t$th most dominated $L$-eigenvector block of $\mathbf{R}$, as expressed by
\begin{equation}\label{chpt:HCE:opt:BlockSelectionUnconstrainedCombiner}
	\mathbf{F}^\mathrm{opt}_t=(\mathbf{U}_{[(t-1)L+1:tL]})^\Hm.
\end{equation}
Evidently, the \emph{Block Selection} method has low complexity, yet leading to non-optimal performance.

In order to improve the performance of channel estimation with multiple trainings, we formulate the optimization problem to design combiners for multiple trainings by recalling (\ref{chpt:HCE:eq:ModifiedGeneralMSE}) shown as follows:
\begin{equation}\label{chpt:HCE:opt:DesignUnconstrainedRFCombinerMultipleTrainings}
	\begin{aligned}
		&\underset{\mathbf{\tilde{V}}_t}{\text{maximize}}\quad{}\mathrm{MSE} \\
		&\text{subject to:}\quad\mathbf{\tilde{V}}_t^\Hm\mathbf{\tilde{V}}_t=\mathbf{I}_L,\quad{} t=1,\cdots,T
	\end{aligned}
\end{equation}
where the $\mathrm{MSE}$ of the objective function is expressed by (\ref{chpt:HCE:eq:ModifiedGeneralMSE}).
The designed $\mathbf{\tilde{V}}_t$ can be employed to construct the RF combiners according to Theorem \ref{chpt:HCE:thm:UnconstrainedOptimalRFCombiner}.
Hence, it is necessary to find the optimizer of (\ref{chpt:HCE:opt:DesignUnconstrainedRFCombinerMultipleTrainings}) to design RF combiners for multiple trainings.
Dealing with the optimization problem, we propose two methods to solve it in the following part of this section.

\subsubsection{Sequential Optimization}
It has been investigated in Subsection \ref{chpt:HCE:sec:OptimalCombinerDesignSinglePilotTraining} that the closed-form expression of the optimal combiner can be obtained in the single-training scenario.
For multiple trainings, however, it is difficult to obtain the global optimal solution directly.
Nevertheless, we propose a sequential approach, namely, the \emph{Sequential Optimization} (short as \emph{Sequential}) method, to minimize MSE step-wisely when $T>1$.
To illustrate the \emph{Sequential Optimization} method, recall (\ref{chpt:HCE:eq:ModifiedGeneralMSE}) and reformulate it in the following form:
\begin{equation}\label{chpt:HCE:eq:ReformulatedMSE}
	\mathrm{MSE} = \tr\left(\left(\bm{\Gamma}_T^{-1} + \rho\mathbf{\tilde{V}}_T\mathbf{\tilde{V}}_T^\Hm\right)^{-1}\right),
\end{equation}
where 
\begin{equation}\label{chpt:HCE:eq:GammaIterativeDefinition}
	\bm{\Gamma}_t^{-1} = \begin{cases}
	\bm{\Gamma}_{t-1}^{-1} + \rho\mathbf{\tilde{V}}_{t-1}\mathbf{\tilde{V}}_{t-1}^\Hm, & t > 1 \\
	\bm{\Lambda}^{-1}, & t = 1.
	\end{cases}
\end{equation}
Following the iterative definition of $\Gamma^{-1}_t$ denoted by (\ref{chpt:HCE:eq:GammaIterativeDefinition}), the MSE can be minimized by the \emph{Sequential Optimization} method as follows.

Firstly, by setting $T=1$ the problem reduces to the single-pilot training case which can be solved as presented in Section \ref{chpt:HCE:sec:OptimalCombinerDesignSinglePilotTraining}.

Subsequently, with $T=2$, the MSE denoted by (\ref{chpt:HCE:eq:ReformulatedMSE}) can be reformulated as
\begin{equation}\label{chpt:HCE:eq:MSESequentialSelection}
	\mathrm{MSE} = \tr\left(\left(\bm{\Gamma}_2^{-1} + \rho\mathbf{\tilde{V}}_2\mathbf{\tilde{V}}_2^\Hm\right)^{-1}\right).
\end{equation}
By applying the same manipulations as in Section \ref{chpt:HCE:sec:OptimalCombinerDesignSinglePilotTraining}, the optimal $\mathbf{\tilde{V}}_2$ to minimize the MSE expressed by (\ref{chpt:HCE:eq:MSESequentialSelection}) can be obtained.
Similar to (\ref{chpt:HCE:eq:UnconstrainedOptimizerTildeV}), the optimal $\mathbf{\tilde{V}}_2$, say $\mathbf{\tilde{V}}^\mathrm{opt}_2$, is constructed by the most dominated $L$ eigen-directions of the pencil $(\bm{\Gamma}_2^2, \bm{\Gamma}_2+\rho^{-1}\mathbf{I}_M)$.
In other words, $\mathbf{\tilde{V}}^\mathrm{opt}_2$ is combined by $L$ generalized eigenvectors corresponding to the $L$ largest generalized eigenvalues of $(\bm{\Gamma}_2^2, \bm{\Gamma}_2+\rho^{-1}\mathbf{I}_M)$.
For ease of exposition, we denote $\bm{\Gamma}_2 = \mathrm{diag}\{\gamma_{1,2}, \cdots, \gamma_{M,2}\}$.
The generalized eigenvalues and eigenvectors of pencil $(\bm{\Gamma}_2^2, \bm{\Gamma}_2+\rho^{-1}\mathbf{I}_M)$ can be obtained as $\bm{\tilde{\gamma}}_2 = \left[\tilde{\gamma}_{1,2}, \cdots, \tilde{\gamma}_{M,2}\right]^\mathrm{T}$ and $\mathbf{I}_M$, respectively, where $\tilde{\gamma}_{j,2}=\gamma_{j,2}^2/(\gamma_{j,2}+\rho^{-1})$ for $j=1,\cdots,M$.
Therefore, 
\begin{equation}\label{chpt:HCE:eq:OptimalVSequentialSelection}
	\mathbf{\tilde{V}}^\mathrm{opt}_2 = \mathbf{I}_{M[j_1,\cdots,j_L]},
\end{equation}
where the indices $[j_1,\cdots,j_L]$ are the first $L$ numbers of $[j_1,\cdots,j_M]$ which follow from the descending order of $\tilde{\gamma}_{i,2}$ shown as $\tilde{\gamma}_{j_1,2}\geq\cdots\geq\tilde{\gamma}_{j_M,2}$.

In the similar manner as $T=2$, we can obtain each $\mathbf{\tilde{V}}^\mathrm{opt}_T$ for any $T>2$ successively according to the generalized eigenvalues and eigenvectors of pencil $(\bm{\Gamma}_T^2, \bm{\Gamma}_T+\rho^{-1}\mathbf{I}_M)$.

Finally, the combiners are derived by applying the definition of $\mathbf{\tilde{V}}_i$ and Lemma \ref{chpt:HCE:lem:UnconstrainedRFCombinerStructure}.
For instance, the optimal $\mathbf{F}_{2,\mathrm{opt}}$ can be expressed by
\begin{equation}\label{chpt:HCE:eq:SequentialDesignResult}
	\mathbf{F}^\mathrm{opt}_2=\mathbf{U}^\mathrm{H}_{[j_1,\cdots,j_L]}.
\end{equation}

By reviewing (\ref{chpt:HCE:eq:GammaIterativeDefinition}) and the \emph{Sequential Optimization} method, a natural question arises here: Whether $\bm{\Gamma}_t$ is invertible or not?
To check this question, we start with considering a two-stage training with $\bm{\Gamma}_1$ and $\bm{\Gamma}_2$.
Note that the known covariance matrix $\mathbf{R}$ is assumed to be symmetric and full-rank\footnote{If $\mathbf{R}$ is rank deficient, the method mentioned below (\ref{chpt:HCE:eq:ChannEstMSE}) can be employed for adjustment.}, thus it is positive definite with all eigenvalues positive.
Hence, $\bm{\Gamma}_1$ is composed of positive diagonal elements which indicates that it is invertible.
By setting $T=2$ in (\ref{chpt:HCE:eq:GammaIterativeDefinition}), we have $\bm{\Gamma}_2^{-1} = \bm{\Gamma}_1^{-1} + \rho\mathbf{\tilde{V}}_1\mathbf{\tilde{V}}_1^\Hm$.
In addition to positive diagonal $\bm{\Gamma}_1$, the optimal $\mathbf{\tilde{V}}_1$ solved in Section \ref{chpt:HCE:sec:OptimalCombinerDesignSinglePilotTraining} is shown to be $\mathbf{I}_{M[1:L]}$.
Hence, it is guaranteed that $\bm{\Gamma}_1^{-1} + \rho\mathbf{\tilde{V}}_1\mathbf{\tilde{V}}_1^\mathrm{H}$ is always diagonal with positive entries for $\rho>0$.
So is $\bm{\Gamma}_2$ invertible.
Using the similar manipulation, one can successively verify that $\bm{\Gamma}_{T-1}$ is composed of positive diagonal entries and hence 
$\bm{\Gamma}_{T}$ is diagonal and invertible for any $T>2$.

\begin{remark}
	In the aforementioned illustrations, each $\mathbf{\tilde{V}}^\mathrm{opt}_t$ ($t=1,\cdots,T$) during the \emph{Sequential Optimization} has the same structure in which they are constructed by column-wise permutations of $\mathbf{I}_M$ which denotes the set of generalized eigenvectors of pencil $(\bm{\Gamma}_t^2, \bm{\Gamma}_t+\rho^{-1}\mathbf{I}_M)$, and the selected columns correspond to the $L$ largest generalized eigenvalues of this pencil.
	Accordingly, each combiner for the multiple trainings, namely $\mathbf{F}^\mathrm{opt}_t$, is the Hermitian of the matrix constructed by the consistent column-wise permutations of $\mathbf{U}$ as the construction of $\mathbf{\tilde{V}}^\mathrm{opt}_t$.
	Note that the eigenvectors of the above pencil always compose an identity matrix due to diagonal $\bm{\Gamma}_t$.
\end{remark}

\begin{remark}
	Inevitably, the optimum $\mathbf{\tilde{V}}^\mathrm{opt}_t$ ($t=1,\cdots,T$) obtained by the \emph{Sequential Optimization} method is not a global minimizer of the MSE in (\ref{chpt:HCE:eq:ModifiedGeneralMSE}).
	Nevertheless, it minimizes the MSE reformulated by (\ref{chpt:HCE:eq:ReformulatedMSE}) for each step of iterations, namely the step-wise minimizer.
	The performance of such sub-optimal combiners will be evaluated by numerical results in Section \ref{chpt:HCE:sec:NumericalSimulationResults}.
\end{remark}

\subsubsection{Alternating Optimization}
For further performance enhancement, we solve (\ref{chpt:HCE:opt:DesignUnconstrainedRFCombinerMultipleTrainings}) via joint optimization over $\mathbf{\tilde{V}}_t$'s.
It in many cases can achieve near optimal performance via the alternating optimization to solve non-convex problems.
More specifically, we consider fixing all the other $\mathbf{\tilde{V}}_t$ ($t\neq j$) while optimizing only a single $\mathbf{\tilde{V}}_j$.
And then, iterations are taken to update each $\mathbf{\tilde{V}}_j$ ($j=1,\cdots,T$) alternatively until convergence.
This solution is named as \emph{Alternating Optimization} (short as \emph{Alternating}) throughout this paper.

For an explicit exposure of alternating optimization, we reformulate the MSE expressed by (\ref{chpt:HCE:eq:ModifiedGeneralMSE}) to separate $\mathbf{\tilde{V}}_j$ from the sum as given by
\begin{equation}\label{chpt:HCE:eq:ReformulatedMSESemiJointSelection}
	\mathrm{MSE}(\mathbf{\tilde{V}}_j) = \tr\left(\left(\mathbf{Q}_j^{-1} + \rho\mathbf{\tilde{V}}_j\mathbf{\tilde{V}}_j^\Hm\right)^{-1}\right), \quad j=1,\cdots,T
\end{equation}
where $\mathbf{Q}_j^{-1} = \bm{\Lambda}^{-1} + \rho\sum_{t=1,t\neq j}^{T}\mathbf{\tilde{V}}_t\mathbf{\tilde{V}}_t^\Hm$.
With given values of $\mathbf{\tilde{V}}_t$ ($t\neq j$), $\mathbf{\tilde{V}}_j$ can be solved in the same manner as Subsection \ref{chpt:HCE:sec:OptimalCombinerDesignSinglePilotTraining} where Lemma \ref{chpt:HCE:lm:BlockGeneralizedRayleighQuotient} is applied.
In this paper, the relative MSE increment of each iteration round, namely, 
\[
	\epsilon_j^{(n)}=\left|\mathrm{MSE}\left(\mathbf{\tilde{V}}^{(n)}_j\right)-\mathrm{MSE}\left(\mathbf{\tilde{V}}^{(n-1)}_j\right)\right|\big/\mathrm{MSE}\left(\mathbf{\tilde{V}}^{(n-1)}_j\right),
\]
will be taken as the error measurement.
The iterations continue until $\epsilon_j^{(n)}$ falls below a prescribed tolerance $\epsilon$ and the last iterate $\mathbf{\tilde{V}}_j^{(n)}$ is taken as the solution.

Finally, the unconstrained near-optimal RF combiners are computed from the solved $\mathbf{\tilde{V}}^\mathrm{opt}_t$ ($t=1,\cdots,T$) by applying Lemma \ref{chpt:HCE:lem:UnconstrainedRFCombinerStructure}.
The alternating optimization method is summarized step-by-step in Algorithm \ref{chpt:HCE:algtm:SemiJointSelection}.
\begin{algorithm} 
	\caption{Alternating Optimization to Design Optimal Unconstrained Combiners for Multiple Trainings}
	\label{chpt:HCE:algtm:SemiJointSelection}
	\begin{algorithmic}[1]
		\INPUT $M$, $T$ and $\mathbf{R}$
		\OUTPUT Optimal unconstrained RF combiners for multiple trainings, $\mathbf{F}^\mathrm{opt}_t$ ($t=1,\cdots,T$)
		\STATE Initialize $n=0$ and $\mathbf{\tilde{V}}^{(0)}_t$ for $t=1,\cdots,T$
		\LOOP
		\STATE Use $\mathbf{\tilde{V}}^{(n)}_t$ ($t\neq j$) to obtain $\mathbf{\tilde{V}}_j^{(n)}$ which minimizes MSE expressed by (\ref{chpt:HCE:eq:ReformulatedMSESemiJointSelection}) for $j=1,\cdots,T$ sequentially and update $\mathbf{\tilde{V}}^{(n)}_j$ with the new solved values
		\IF{$\frac{\left|\mathrm{MSE}\left(\mathbf{\tilde{V}}^{(n)}_j\right)-\mathrm{MSE}\left(\mathbf{\tilde{V}}^{(n-1)}_j\right)\right|}{\mathrm{MSE}\left(\mathbf{\tilde{V}}^{(n-1)}_j\right)}<\epsilon$}
		\STATE Stop \textbf{loop}
		\ELSE
		\STATE Update $n$ by $1$ (i.e., $n=n+1$)
		\ENDIF
		\ENDLOOP
		\STATE Compute $\mathbf{F}^\mathrm{opt}_t$ with $\mathbf{\tilde{V}}^\mathrm{opt}_t$ by applying Lemma \ref{chpt:HCE:lem:UnconstrainedRFCombinerStructure}
		\STATE Output $\mathbf{F}^\mathrm{opt}_t$
	\end{algorithmic}
\end{algorithm}

Since Algorithm \ref{chpt:HCE:algtm:SemiJointSelection} is implemented via iterations, the convergence issue is needed to be addressed here.
During each iteration, the solution $\mathbf{\tilde{V}}_j^{(n)}$ always minimizes (\ref{chpt:HCE:eq:ReformulatedMSESemiJointSelection}).
Hence, the objective MSE is monotonically decreasing as the increasing of iterations until it achieves the minimal MSE.
This proves the convergence of Algorithm \ref{chpt:HCE:algtm:SemiJointSelection}.

\subsection{Design of Phase-only RF Combiners}\label{chpt:HCE:sec:DesignOfPhaseOnlyCombiners}
In the previous subsections, we first omit the phase-only constraints when designing the RF combiners.
However, the RF combiners are implemented using phase shifters in practical hybrid systems, which can only perform phase rotations on the received RF signals.
Hence, the magnitude of each element of a RF combiner matrix must keep constant.
In this subsection, we propose a method to design the phase-only RF combiners based on the unconstrained optimal ones.
Moreover, the performance loss incurred by the constant-magnitude constraint is characterized by numerical evaluation shown in Section \ref{chpt:HCE:sec:PerformanceEvaluationOfPhaseOnlyRFCombiners}.

Inspired by the fact that the phase shifters only perform phase adjustments on the received signal, we take out the phase of each unconstrained combiner and construct the constant-magnitude RF combiners according to 
\begin{equation}\label{chpt:HCE:eq:ConstrainedCombinerDesignDirectPhase}
	\mathbf{F}_t^{(i,j)} = e^{\mathrm{j}\phi_t^{(i,j)}}, t = 1,\cdots,T,
\end{equation}
where $\phi_t^{(i,j)}$ denotes the phase of the ($i$,$j$)th element of $\mathbf{F}_{t,\mathrm{opt}}$, i.e., $\mathbf{F}_{t,\mathrm{opt}}^{(i,j)}=a_t^{(i,j)}e^{\mathrm{j}\phi_t^{(i,j)}}$.

It is interesting to find that such heuristic phase-only RF combiners achieve desirable performance via simulation verifications in Section \ref{chpt:HCE:sec:NumericalSimulationResults}.
Besides its effectiveness, the very low complexity is another advantage of this design.
Furthermore, the performance of the phase-only combiners designed above will be examined and analyzed in Subsections \ref{chpt:HCE:sec:PerformanceEvaluationOfPhaseOnlyRFCombiners} and \ref{chpt:HCE:sec:SpectralEfficiencyEvaluationWithHybridChannelEstimations} in terms of both MSE of channel estimation and spectral efficiency in hybrid precoded multiuser massive MIMO communications.

\section{Estimation of Spatial Channel Covariance}\label{chpt:HCE:sec:EstimationOfSpatialChannelCovariance}
The prior knowledge of spatial channel correlations, which can be characterized by the channel covariance matrix, is one of the most important prerequisites to improve the performance of not only channel estimations but also precoding designs \cite{Liu2015Two}.
While it has been assumed to be known by the BS in our proposed channel estimation framework, the covariance estimation should not be overlooked in practice.
Due to the limited RF chains, it is infeasible to estimate the covariance matrix in hybrid precoding systems by employing conventional approaches \cite{Carlson1988Covariance,Liang2001Downlink,Melvin2006approach,Mendez-Rial2015Augmented}.
Dealing with this problem, some of the prior works have made their efforts, but still have limitations when applied to practical hybrid-structured massive MIMO or mmWave systems.
\cite{Chi2013PETRELS} has proposed a parallel subspace estimation and tracking by recursive least squares from partial observations, namely the PETRELS method, which can be employed in the hybrid massive MIMO system to perform covariance estimation, whereas tremendous observations are needed to achieve desirable performance due to no consideration on the symmetry property of a spatial channel covariance matrix.
A less complicated method has been proposed in \cite{Mendez-Rial2015Adaptive}, however, by employing a parallel switch network which increases the hardware cost.
In \cite{Park2016Spatial}, a compressive sensing based spatial channel covariance estimation has been proposed while a strong condition is assumed that AoAs do not change during the long-term covariance estimations which is impractical.
In this section, we propose a practical covariance matrix estimation method which can be employed in a hybrid-structured massive MIMO system with limited RF chains.
The estimates of spatial channel covariance matrix can be utilized in our proposed channel estimation framework.

Let $T=M/L$ and consider the $i$th ($i=1,\cdots,N_\mathrm{c}$) coherence interval, (\ref{chpt:HCE:eq:CompactInputOutput}) can be rewritten as
\begin{equation}\label{chpt:HCE:eq:CompactInputOutputIntervalI}
	\mathbf{y}_\mathrm{c}[i] = \sqrt{\rho}\mathbf{F}_\mathrm{c}[i]\mathbf{g}[i] + \mathbf{F}_\mathrm{d}[i]\mathbf{n}_\mathrm{c}[i],
\end{equation}
where $\mathbf{F}_\mathrm{c}[i]\in\CX^{M\times M}$ denotes the compacted phase shifting matrix in the $i$th coherence interval, $\mathbf{F}_\mathrm{d}[i]\triangleq[\mathbf{F}^\mathrm{T}_1[i],\cdots,\mathbf{F}^\mathrm{T}_T[i]]^\mathrm{T}$, $\mathbf{n}_\mathrm{c}[i]\in\CX^{ML\times 1}$ is the AWGN and $N_\mathrm{c}$ stands for the number of coherence intervals during which the spatial channel covariance matrix can be viewed constant, i.e., $\mathbf{R}=\E\left\{\mathbf{g}[i]\mathbf{g}[i]^\mathrm{H}\right\}$.
For the purpose of simplifying trainings, it is durable to employ the same phase shifting matrix for each coherence interval, i.e., $\mathbf{F}_\mathrm{c}[i]=\mathbf{F}_\mathrm{c}$.
Further, we can select an invertible $\mathbf{F}_\mathrm{c}$ from the feasible constant-magnitude matrix set $\mathcal{F}$ for training purpose, e.g., the DFT matrix.
Hence, the spatial covariance matrix of the channel can be derived from (\ref{chpt:HCE:eq:CompactInputOutputIntervalI}) as
\begin{equation}\label{chpt:HCE:eq:GeneralCovarianceMatrix}
	\mathbf{R} = \frac{1}{\rho}\mathbf{F}_\mathrm{c}^{-1}\left(\mathbf{R}_{\mathbf{y}_\mathrm{c}}-\mathbf{F}_\mathrm{d}\mathbf{F}_\mathrm{d}^\mathrm{H}\right)\left(\mathbf{F}_\mathrm{c}^{-1}\right)^\mathrm{H},
\end{equation}
where $\mathbf{R}_{\mathbf{y}_\mathrm{c}}=\E\left\{\mathbf{y}_c[i]\mathbf{y}_c[i]^\mathrm{H}\right\}$ denoting the covariance matrix of the received signal.

From (\ref{chpt:HCE:eq:GeneralCovarianceMatrix}), the estimation of spatial channel covariance matrix can be converted to that of the received signal covariance, where both phase shifting matrix $\mathbf{F}_\mathrm{c}$ and pilot transmission power $\rho$ are assumed to be known by the BS.
In practice, the statistical covariance of the received signal can be approximated with the sample covariance matrix constructed as follows:
\begin{equation}\label{chpt:HCE:eq:SampleCovarianceMatrixReceivedSignal}
	\mathbf{\hat{R}}_{\mathbf{y}_\mathrm{c}}=\frac{1}{N_\mathrm{c}}\sum_{i=1}^{N_\mathrm{c}}\mathbf{y}_\mathrm{c}[i]\mathbf{y}_\mathrm{c}[i]^\mathrm{H}.
\end{equation}
By substituting (\ref{chpt:HCE:eq:SampleCovarianceMatrixReceivedSignal}) into (\ref{chpt:HCE:eq:GeneralCovarianceMatrix}), it derives the estimate of the spatial channel covariance matrix.

Note that it may need many trainings to approximate the covariance matrix of the received signal with its sample covariance, which increases the overhead of channel estimation.
One may concern how much resource is consumed to estimate channel covariance in practical systems.
Theoretically, the covariance matrix exploits the second-order statistics of the Wide-Sense Stationary Uncorrelated Scattering (WSSUS) channel nature which varies slowly over time and frequency, and a one-time estimate can be used for long-term communications \cite{Haghighatshoar2017Massive}.
In fact, the experimental results in \cite{Viering2002Spatial} have revealed that the spatial covariance matrix can be viewed constant over 22.6 s, 9.7 s and 4.8 s in the urban, rural and indoor environment, respectively.
Considering a system with 10 MSps, for example, the estimated spatial channel covariance can be used for $1\times10^9$ times within 10 s, which is far more than the consumption of channel usage on covariance estimation.
Therefore, the overhead of covariance estimation is negligible in practice.	
Apart from overhead, the estimation accuracy of covariance matrix and its impact on channel estimation performance is another issue need to be considered in practical systems.
According to (\ref{chpt:HCE:eq:GeneralCovarianceMatrix}) and (\ref{chpt:HCE:eq:SampleCovarianceMatrixReceivedSignal}), the estimation accuracy of $\mathbf{R}$ is determined by $\mathbf{\hat{R}}_{\mathbf{y}_\mathrm{c}}$, which can be increased by extending the averaging time.
In Section \ref{chpt:HCE:sec:NumericalSimulationResults}, simulation results will demonstrate the number of channel usages needed to achieve the required estimation accuracy.

\section{Numerical and Simulation Results}\label{chpt:HCE:sec:NumericalSimulationResults}
In this section, we evaluate the performance of the proposed channel estimation scheme by employing both unconstrained and phase-only RF combiners to perform uplink channel estimation in the massive MIMO system with the hybrid precoding structure.
The evaluations are performed with nonparametric channel models followed by parametric ones.

In simulations, we consider both single-user and multi-user scenarios.
In the single-user scenario, a single-antenna MS is served by a BS deployed in the hybrid analog/digital system structure which is shown by Fig. \ref{chpt:HCE:fig:SystemModel} with $K=1$.
The multi-user scenario adopts $K=8$ single-antenna MSs served by the BS with the same structure as that employed in the single-user case.
Without explicit indications, the BS is equipped with $M=64$ antennas followed by $L=8$ RF chains.
As the power of noise is normalized to $1$, the transmission power $\rho$ can be used to denote the received signal-to-noise ratio (SNR).
Each simulation result is obtained by averaging over $10,000$ channel realizations.
Moreover, we adopt the low-complexity hybrid precoding scheme proposed by Liang \emph{et al.} in \cite{Liang2014Low} to pre-process the transmission signal in downlink communications.

\subsection{Performance with Nonparametric Channel Model}
In this subsection, the nonparametric channel model denoted by $\mathbf{g} = \mathbf{R}^{1/2}\mathbf{h}$ is considered, where $\mathbf{h}\in\CX^{M\times1}$ stands for the small-scale fading of the channel constructed by independently and identically distributed (i.i.d.) Gaussian random variables, i.e., $\mathbf{h}\sim\CN(\mathbf{0}, \mathbf{I}_M)$, and $\mathbf{R}$ is modeled by exponential correlation model as $[\mathbf{R}]_{m,n} = a^{|m-n|}$ with $[\mathbf{R}]_{m,n}$ denoting the $(m,n)$th element of $\mathbf{R}$ and $0\leq a < 1$ being a real number that controls the correlation introduced to the channel model \cite{Choi2014Downlink,Biguesh2006Training}.
Here, a larger $a$ corresponds to a more spatially correlated channel and $\mathbf{g}$ degenerates to an i.i.d. Rayleigh fading channel when $a=0$.

\subsubsection{Single Training Performance Evaluation}
\label{chpt:HCE:sec:SingleTrainingPerformanceEvaluation}
\begin{figure}[htbp]
	\centering
	\includegraphics[width=\linewidth]{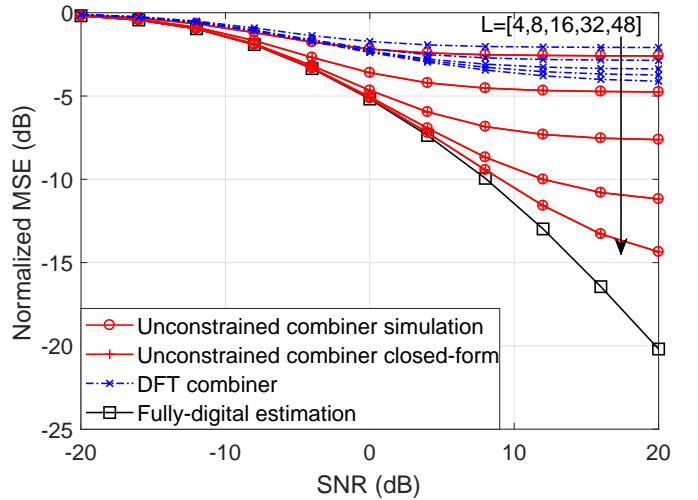}
	\caption{Performance comparison of the single-training hybrid channel estimation with different RF chains. ($M=64$, $a=0.8$)}
	\label{chpt:HCE:fig:ChanEstVsSnrSingleTraining}
\end{figure}
Fig. \ref{chpt:HCE:fig:ChanEstVsSnrSingleTraining} shows the normalized MSE performance of the hybrid channel estimations with the designed single-training unconstrained combiner at different numbers of RF chains.
As a benchmark, the performance of the fully-digital estimation is also presented.
Note that the fully-digital estimation is performed with single training throughout this section.
From the comparison, it is evident to find that the MSE decreases with the increase of the pilot training power, which verifies Corollary \ref{chpt:HCE:cry:UnconstrainedOptimizerRFCombiner3}.
Moreover, with more RF chains equipped at the BS, the channel estimation performance is also improved, which confirms Corollary \ref{chpt:HCE:cry:UnconstrainedOptimizerRFCombiner2}.
Note that in Fig. \ref{chpt:HCE:fig:ChanEstVsSnrSingleTraining}, we also plot the estimation performance with RF combiners composed of the columns randomly selected from a discrete Fourier transform (DFT) matrix.
Its poor performance as compared to the designed combiners proposed in this paper justifies the necessity to properly design RF combiners and the effectiveness of our design.
Furthermore, it is observed that the derived closed-form expressions are quite accurate in characterizing the MSE performance of the channel estimations throughout the whole SNR range, thus providing valuable guidelines in practical system designs.
Finally, the figure also shows that the performance gap between the fully-digital estimation and the limited-chain hybrid channel estimations is tolerably small at low SNR regions.
There is only about $0.5$ dB MSE gap between the hybrid 16-chain and the conventional fully-digital estimations at SNR=$0$ dB, where three-quarter RF chains are saved with the hybrid structure.
However, one may argue: how about the performance gap at high SNR regions?
Due to limited chains, the single training is not sufficient to achieve the performance of fully-digital channel estimation.
To improve the performance of the hybrid channel estimations at full SNR regions, multiple trainings are introduced in this paper and the performance is examined in the next subsection.

\subsubsection{Multiple Trainings Performance Evaluation}
\label{chpt:HCE:sec:MultipleTrainingsPerformanceEvaluation}
\begin{figure}[htbp]
	\centering
	\includegraphics[width=\linewidth]{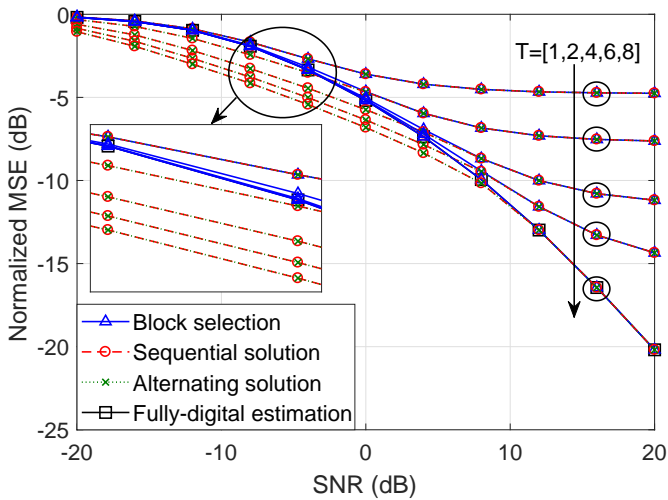}
	\caption{Performance comparison of the multiple-training hybrid channel estimation with different training times. ($M=64$, $L=8$, $a=0.8$)}
	\label{chpt:HCE:fig:ChanEstVsSnrMultipleTraining}
\end{figure}
In this paper, the design of unconstrained combiners for multiple trainings is formulated as (\ref{chpt:HCE:opt:DesignUnconstrainedRFCombinerMultipleTrainings}), which are solved by both \emph{Sequential} and \emph{Alternating} approaches alongside of the intuitive \emph{Block Selection} method.
The performance is examined in Fig. \ref{chpt:HCE:fig:ChanEstVsSnrMultipleTraining} with $T=1,2,4,6$ and $8$.
The figure shows that the normalized MSE of the hybrid channel estimation decreases as the increasing of training times due to the increased DoF of baseband observations.
Additionally, it is interesting to find that the MSE performance with $L=8$ and $T=8$ almost achieves that of the fully-digital estimation at high SNRs.
This phenomenon verifies that the performance of fully-digital channel estimation can be achieved by the limited-chain hybrid channel estimation with sufficient DoF of trainings, i.e., $T\times L=M$.
In the low SNR regions, it is evident to find that even $T=2$ trainings can outperform the fully-digital estimation.
This phenomenon reveals the fact that, when the dimension of the dominated subspaces of the correlated channels is far less than $M$, the limited-chain estimation can suppress noise better than the fully-digital one due to multiple trainings, which results in the distinct performance gap in the low SNR regions.
On top of that, more performance differences can also be compared from the observation of Fig. \ref{chpt:HCE:fig:ChanEstVsSnrMultipleTraining}.
In the figure, the normalized MSE of the \emph{Block Selection} method denoted by the triangle curve is always larger than or equal to that of the other two methods, i.e., the \emph{Sequential} and \emph{Alternating} solutions.
The performance gap is particularly obvious at the low SNR regions.
Such performance loss of \emph{Block Selection} is caused by the fact that the noise is not considered when designing the combiners for the second and later pilot trainings.
However, little performance gap can be observed between \emph{Block Selection} and the others at high SNR regions due to the less significance of noise to channel estimation.
By comparing the performances between the \emph{Sequential} and \emph{Alternating} solutions, it is evidenced that they achieve almost the same MSE no matter at high or low SNR regions which implies that the \emph{Sequential} method can achieve the local optimum in solving (\ref{chpt:HCE:opt:DesignUnconstrainedRFCombinerMultipleTrainings}).
On the other hand, the lower designing complexity of the \emph{Sequential} designates its superiority than the \emph{Alternating} method.

\subsubsection{Performance Evaluation of Phase-only RF Combiners}\label{chpt:HCE:sec:PerformanceEvaluationOfPhaseOnlyRFCombiners}
\begin{figure}[htbp]
	\centering
	\includegraphics[width=\linewidth]{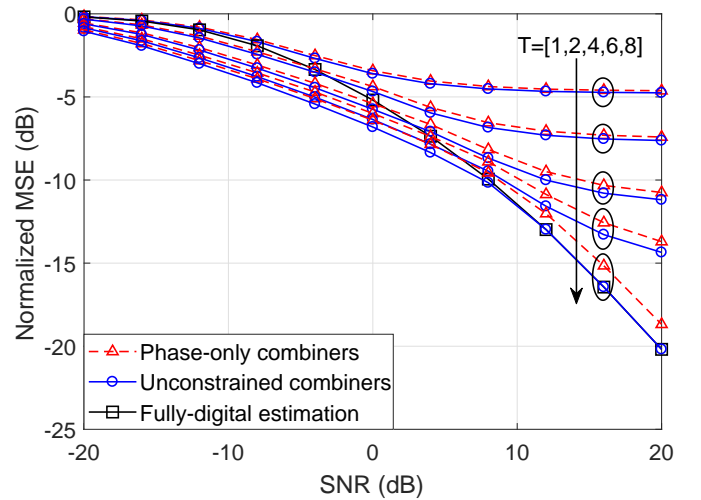}
	\caption{Performance comparison of channel estimation between unconstrained and phase-only combiners with the \emph{Sequential} method under different pilot trainings. ($M=64$, $L=8$, $a=0.8$)}
	\label{chpt:HCE:fig:ChanEstVsSnrMultipleTrainingRFvsUnc}
\end{figure}
\begin{figure}[htbp]
	\centering
	\includegraphics[width=\linewidth]{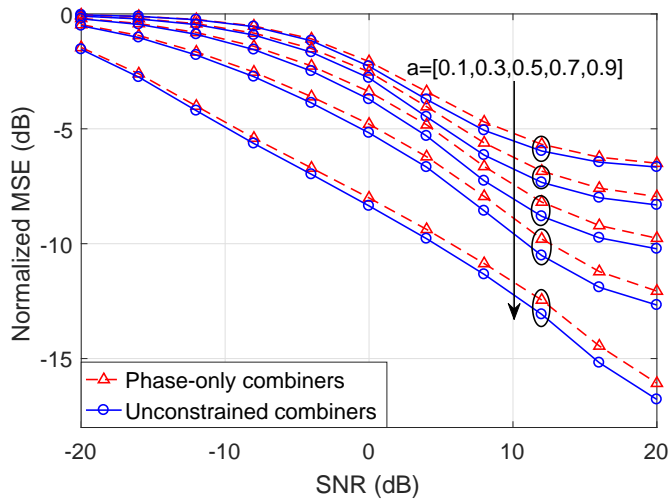}
	\caption{Channel estimation performance of both unconstrained and phase-only combiners with the \emph{Sequential} method under different channel correlations. ($M=64$, $L=8$, $T=6$)}
	\label{chpt:HCE:fig:ChanEstVsCorrMultipleTraining}
\end{figure}

In the previous two subsections, the performance of the hybrid channel estimation with unconstrained RF combiners are evaluated.
However, only phase shifters can be employed in the current practical applications with hybrid precoding structure, which means that only phase-only combiners are applicable in practice.
In this subsection, we examine the performance in hybrid channel estimation with the phase-only combiners and compare it to that of the corresponding unconstrained ones, where the combiners for multiple trainings are designed by employing the \emph{Sequential} method.
In Fig. \ref{chpt:HCE:fig:ChanEstVsSnrMultipleTrainingRFvsUnc}, the dashed lines marked by triangles denote the MSE performance of the hybrid channel estimation with phase-only combiners which are designed according to (\ref{chpt:HCE:eq:ConstrainedCombinerDesignDirectPhase}), where the performance of the unconstrained combiners are denoted by the circled solid lines.
From comparisons, it is obvious that the performance loss caused by the phase-only combiners is tolerable at high SNR regions and is negligible throughout the low SNR regions.
For instance, there is only $1.5$ dB MSE loss at SNR=$20$ dB while $0.08$ dB loss at SNR=$-20$ dB when $T=8$.
The figure also shows that the performance loss is smaller with fewer trainings, i.e., smaller $T$.
Therefore, the hybrid channel estimation with phase-only combiners achieve the desirable performance measured by the normalized MSE.
Furthermore, we also present the performance of hybrid channel estimation at different channel correlation shown in Fig. \ref{chpt:HCE:fig:ChanEstVsCorrMultipleTraining}.
It is evident that the normalized MSE of the hybrid channel estimation decreases, thus the performance increases, as the increasing of $a$, namely, with more spatially correlated channels, which verifies Corollary \ref{chpt:HCE:cry:UnconstrainedOptimizerRFCombiner1}.
Moreover, it is obvious that the performance between the phase-only and unconstrained combiners is tight throughout the SNR regions, e.g., there exist only $0.7$ dB of MSE gap at SNR=$20$ dB when $a=0.9$.

\subsubsection{Spectral Efficiency Evaluation with Hybrid Channel Estimation}
\label{chpt:HCE:sec:SpectralEfficiencyEvaluationWithHybridChannelEstimations}
\begin{figure}[htbp]
	\centering
	\includegraphics[width=\linewidth]{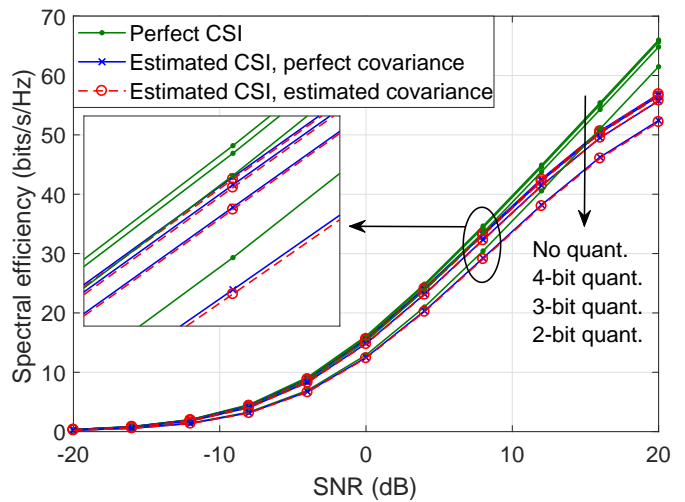}
	\caption{Spectral efficiency performance with the hybrid precoding scheme using both estimated and perfect CSI where the non-perfect CSI is estimated with both perfect and estimated spatial channel covariance matrix. ($M=64$, $L=8$, $K=8$, $T=8$, $N_\mathrm{c}=300$, $\rho=10$ dB and $a=0.8$)}
	\label{chpt:HCE:fig:SpectralEfficiencyWithEstR}
\end{figure}

Apart from the normalized MSE performance presented in previous subsections, the spectral efficiency of the hybrid channel estimation is examined in this part.
In the simulation, the spatial channel covariance matrix of each user is estimated by the method proposed in Section \ref{chpt:HCE:sec:EstimationOfSpatialChannelCovariance} with $300$ training intervals.
The following channel estimation is performed with the estimated covariance matrix.
Within each coherence interval, $K$ MSs transmit orthogonal training pilots to the BS for channel estimations during uplink communications, while the BS broadcasts data to all MSs during the downlink communications by employing the hybrid precoding with the estimated channels.
Note that the uplink and downlink channels are assumed reciprocal in this simulation and the spectral efficiency is calculated over the downlink data transmissions.
Fig. \ref{chpt:HCE:fig:SpectralEfficiencyWithEstR} shows comparisons of the spectral efficiency with the precoding scheme designed from the perfect CSI and estimated CSI with perfect and estimated covariance matrices, respectively, where the phase shifters are considered both ideal with no quantizations and non-ideal with 2-bit, 3-bit and 4-bit quantizations, respectively.
From the figure, we observe that the spectral efficiency curves denoting the estimated and ideal covariance matrices are very tight which means that the estimated covariance matrix with 300 training intervals is accurate enough to be comparable with the ideal one when employed in hybrid channel estimations.
By recalling the example in Section \ref{chpt:HCE:sec:EstimationOfSpatialChannelCovariance}, it is shown that the overhead of covariance estimation is only $2.4\times10^{-6}$ per user, which is negligible in practice.
The figure further shows that the spectral efficiency with the estimated CSI approaches that with the perfect one, especially in low SNR region which is always the scenario in realistic massive MIMO systems.
Moreover, we can see that the hybrid precoding scheme with estimated CSI achieves highly desirable performance as compared to the gene-assisted system where perfect CSI is available to the transmitter, particularly in the low SNR regions, where the negligible performance gap is introduced by the constant-magnitude constraint.
Finally, the curves denoting different phase quantizations show that the 2-bit quantization is not sufficient to achieve the no-quantization performance while the 3-bit, especially the 4-bit, quantized phase shifters produce comparable spectral efficiency to the no-quantized ideal phase shifters, with only slight performance gap, which means that our proposed hybrid channel estimation is insensitive to the quantization of phase shifters.

\subsection{Performance with Spatial Channel Model}
\label{chpt:HCE:sec:PerformanceEvaluationWithParametricChannelModel}
\begin{figure}[htbp]
	\centering
	\includegraphics[width=\linewidth]{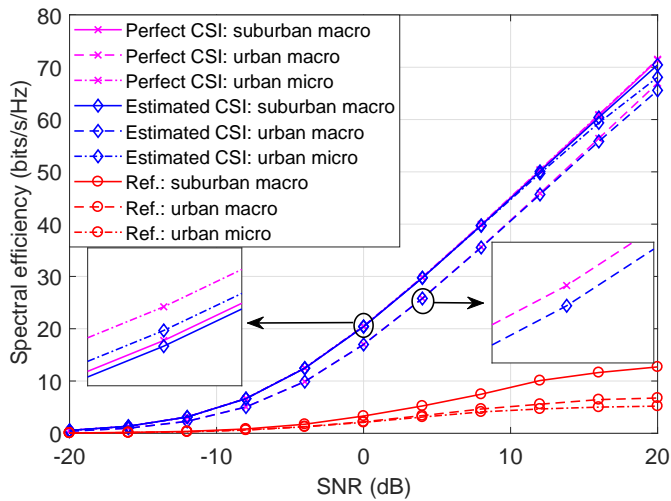}
	\caption{Spectral efficiency performance with the hybrid precoding scheme in practical SCM channels using both perfect and estimated CSI. The Ref. result follows the channel estimation algorithm proposed in \cite{Alkhateeb2014Channel}. ($M=64$, $L=8$, $T=6$, $N_\mathrm{c}=1000$, $\rho=20$ dB and 3-bit phase quantization)}
	\label{chpt:HCE:fig:SpectralEfficiencyScm}
\end{figure}

In Subsection A, the nonparametric channel models are employed in simulations to evaluate the performance of the proposed channel estimation design in the hybrid precoding system.
In this subsection, the proposed channel estimation framework is evaluated in the practical spatial channel model (SCM) under suburban macro, urban macro and urban micro environments.
The SCM channels are generated from the open source software provided by \cite{Salo2005MATLAB} with default settings (6 scattering paths with 20 subpaths each, which is rich scattering) and random seed 1.
The spatial channel covariance matrix is estimated by the proposed method in Section \ref{chpt:HCE:sec:EstimationOfSpatialChannelCovariance}.
Fig. \ref{chpt:HCE:fig:SpectralEfficiencyScm} shows the spectral efficiency performances of the perfect and estimated CSI obtained by the proposed method and by the compressed sensing based channel estimation in \cite{Alkhateeb2014Channel}, respectively.
From the figure, it is obvious that the spectral efficiency with the proposed channel estimation method almost overlaps that with the perfect CSI in all three environments with only visible minor performance gap in the high SNR region, i.e., 15 to 20 dB, which explicitly show the high accuracy of the proposed channel estimation frame.
Note that only 6 trainings are employed in the simulation which verifies that less than $M/L$ trainings can be sufficient to achieve the performance of the perfect CSI in spatially correlated channels.
Moreover, the comparisons with the compressed sensing based hybrid channel estimations proposed in \cite{Alkhateeb2014Channel} evidently show that our proposed method is more effective in the rich scattering channels which verifies that no sparsity is required by our proposed method to estimate CSI.

\section{Conclusions}\label{chpt:HCE:sec:Conclusions}
This paper has investigated and proposed a channel estimation framework for massive MIMO systems employing a hybrid transceiver structure with a limited number of RF chains and a phase shifting network.
The optimal RF combiners have been designed in the case of single training by exploiting the channel covariance matrix and the complete CSI is estimated in the hybrid structured massive MIMO system.
The theoretical analyses have shown that the channel estimation performance can improve when the channel is more spatially correlated, the BS deploys more RF chains or the MSs increase training power, which has been verified by simulation results.
In addition, the RF combiners have been designed in the case of multiple trainings to increase the DoF of the received signal measurements at BS.
The simulation results reveal that multiple trainings can approach the performance of fully-digital estimation with single training and even outperform it at low SNR regions when the channel is highly correlated.
Finally, this paper has proposed a practical spatial channel covariance matrix estimation method in the hybrid-structured massive MIMO system.
The simulation results have shown the effectiveness of the proposed method under various scenarios in terms of the spectral efficiency with the hybrid precoding.

\begin{appendix}
\subsection{Proof of Lemma \ref{chpt:HCE:lem:UnconstrainedRFCombinerStructure}}\label{chpt:HCE:prf:lem:UnconstrainedRFCombinerStructure}
By applying the eigenvalue decomposition of a positive-definite $\mathbf{R}$, i.e., $\mathbf{R}=\mathbf{U}\bm{\Lambda}\mathbf{U}^\mathrm{H}$, the objective function in (\ref{chpt:HCE:opt:DesignPhaseShifter}) can be derived as
\begin{equation}\label{chpt:HCE:prf:UnconstrainedRFCombinerStructure:eq:1}
	\begin{aligned}
		\mathrm{MSE} =& \tr\left(\left(\mathbf{R}^{-1}+\rho\mathbf{F}_\mathrm{c}^\Hm\mathbf{R}_{\mathbf{F}_\mathrm{d}}^{-1}\mathbf{F}_\mathrm{c}\right)^{-1}\right) \\
		=& \tr\left(\left(\bm{\Lambda}^{-1}+\rho\mathbf{U}^\Hm\mathbf{F}_\mathrm{c}^\Hm\mathbf{R}_{\mathbf{F}_\mathrm{d}}^{-1}\mathbf{F}_\mathrm{c}\mathbf{U}\right)^{-1}\right)  \\
		=&\tr\left(\left(\bm{\Lambda}^{-1}+\rho\mathbf{U}^\Hm\left(\sum_{i=1}^{p}\mathbf{F}_i^\Hm(\mathbf{F}_i\mathbf{F}_i^\Hm)^{-1}\mathbf{F}_i\right)\mathbf{U}\right)^{-1}\right),
	\end{aligned}
\end{equation}
where the last equality uses 
\begin{equation}
	\mathbf{R}_{\mathbf{F}_\mathrm{d}}^{-1} = \mathrm{blkdiag}\left\{(\mathbf{F}_1\mathbf{F}_1^\Hm)^{-1}, \cdots, (\mathbf{F}_T\mathbf{F}_T^\Hm)^{-1}\right\}.
\end{equation}
Substituting (\ref{chpt:HCE:eq:UnconstrainedRFCombinerSVD}) into the above equation yields
\begin{equation}
	\mathrm{MSE} = \tr\left(\left(\bm{\Lambda}^{-1}+\rho\sum_{i=1}^{T}\mathbf{U}^\Hm\mathbf{V}_{i,L}\mathbf{V}_{i,L}^\Hm\mathbf{U}\right)^{-1}\right),
\end{equation}
which indicates that the MSE is independent of both $\mathbf{U}_i$ and $\bm{\Sigma}_i$.
Hence, we can safely set both of them to identity matrices and obtain $\mathbf{F}^\mathrm{opt}_i = \mathbf{V}_{i,\mathrm{L}}^\mathrm{H}$ from (\ref{chpt:HCE:eq:UnconstrainedRFCombinerSVD}) without loss of the optimality, which completes the proof.

\subsection{Block Generalized Rayleigh Quotient}\label{chpt:HCE:apdx:BlockGeneralizedRayleighQuotient}
\begin{lemma}\label{chpt:HCE:lm:BlockGeneralizedRayleighQuotient}
	(Block Generalized Rayleigh Quotient)\cite{Baker2008Riemannian} Given a full column rank matrix $\mathbf{V}$, the block generalized Rayleigh quotient with respect to the pencil ($\mathbf{A}$, $\mathbf{B}$) is defined as
	\begin{equation}\label{chpt:HCE:eq:BlockGeneralizedRayleighQuotient}
		\mathrm{GRQ}(\mathbf{V}) = \tr\left(\left(\mathbf{V}^\T\mathbf{B}\mathbf{V}\right)^{-1}\mathbf{V}^\T\mathbf{A}\mathbf{V}\right),
	\end{equation}
	where $\mathbf{V}\in\CX^{N\times T}$.
	Suppose the generalized eigenvalues and the corresponding eigenvectors of the pencil ($\mathbf{A}$, $\mathbf{B}$) are denoted by $\lambda_1\geq\cdots\geq\lambda_N$ and $(\mathbf{v}_1,\cdots,\mathbf{v}_N)$, respectively, where the eigenvalues are arranged in decreasing order, without loss of generality.
	For a matrix containing a subset of distinct generalized eigenvectors, the generalized Rayleigh quotient evaluates to the sum of the associated eigenvalues:
	\begin{equation}\label{chpt:HCE:eq:BlockGeneralizedRayleighQuotientValue}
		\mathrm{GRQ}([\mathbf{v}_{i_1}, \cdots, \mathbf{v}_{i_T}]) = \sum_{j=1}^{T}\lambda_{i_j}
	\end{equation}
	and bounds can be placed on the Rayleigh quotient of an $N\times T$ matrix as
	\begin{equation}\label{chpt:HCE:eq:BlockGeneralizedRayleighQuotientBounds}
		\lambda_1+\lambda_2+\cdots+\lambda_T \geq \mathrm{GRQ}(\mathbf{V}) \geq \lambda_{N-T+1} + \cdots + \lambda_N,
	\end{equation}
	where an equality in this bound indicates that $\mathbf{V}$ is composed of the extreme eigenvectors as
	$\mathrm{colsp}(\mathbf{V}) = \mathrm{colsp}([\mathbf{v}_1, \cdots, \mathbf{v}_T])$ or $\mathrm{colsp}(\mathbf{V}) = \mathrm{colsp}([\mathbf{v}_{N-T+1}, \cdots, \mathbf{v}_N])$.
	Here $\mathrm{colsp}(\cdot)$ represents the column space spanned by the specified matrix.
\end{lemma}

\subsection{Proof of Corollary \ref{chpt:HCE:cry:UnconstrainedOptimizerRFCombiner1}}\label{chpt:HCE:prf:cry:UnconstrainedOptimizerRFCombiner1}
We start with the definition of function $\phi:\RL^L\rightarrow\RL$ as
\begin{equation}
	\phi(\mathbf{x}) = \sum_{l=1}^{L}\psi(x_l),
\end{equation}
where $\psi(x_l)\triangleq\frac{x_l^2}{x_l+1/\rho}$.
Note that the MSE shown in (\ref{chpt:HCE:eq:MSEOfSingleTraining}) can be denoted by $\mathrm{MSE}=M-\phi(\bm{\lambda})$.
It is straightforward to show that $\psi(x_l)$ is convex and hence $\phi(\mathbf{x})$ is \emph{Schur-convex} according to \cite{Marshal2010Inequalities}.
Suppose channel $\mathbf{g}_1\in\CX^M$ is \emph{more spatially correlated} than $\mathbf{g}_2\in\CX^M$ which can be mathematically expressed as $\bm{\lambda}_1\succ\bm{\lambda}_2$, where $\bm{\lambda}_1=[\lambda_{1,1}, \cdots, \lambda_{M,1}]^\T$ and $\bm{\lambda}_2=[\lambda_{1,2}, \cdots, \lambda_{M,2}]^\T$ denote the eigenvalues sorted in descending order of the covariance matrices of channel $\mathbf{g}_1$ and $\mathbf{g}_2$, respectively.
Applying the majorization theory on the Schur-convex function $\phi(\mathbf{x})$, there exists
\begin{equation}
	\phi(\bm{\lambda}_1) \geq \phi(\bm{\lambda}_2).
\end{equation}
Hence, $\mathrm{MSE}(\mathbf{g}_1) \leq\mathrm{MSE}(\mathbf{g}_2)$, which proves Corollary \ref{chpt:HCE:cry:UnconstrainedOptimizerRFCombiner1}.
\end{appendix}


\bibliographystyle{IEEEtran}
\bibliography{main}

\end{document}